\documentstyle[prd,aps]{revtex}

\newcommand{\beq}{\begin{equation}}
\newcommand{\eeq}[1]{\label{#1}\end{equation}}
\newcommand{\beqn}{\begin{eqnarray}}
\newcommand{\eeqn}[1]{\label{#1}\end{eqnarray}}
\newcommand{\ba}{\begin{array}}
\newcommand{\ea}{\end{array}}

\newcommand{\D}{{\cal{D}}}

\begin{document}
\draft
\title{Schwinger terms in  Weyl-invariant and diffeomorphism-invariant 2-d 
scalar field theory\thanks
{This work is supported in part by a Schr\"odinger Stipendium of the Austrian
FWF, an External Scholarship from CONICET, Argentina, and funds provided by
the U.S.\ Department of Energy (D.O.E.) under cooperative research
agreement \#DF-FC02-94ER40818.}}
\author{Christoph Adam$^{a,b}$
\thanks{E-mail address: {\tt adam@ctp.mit.edu, adam@pap.univie.ac.at}} 
and 
Gerardo L. Rossini$^{a,c}$
\thanks{E-mail address: {\tt rossini@ctp.mit.edu, 
rossini@venus.fisica.unlp.edu.ar}}
\\[3mm]
{\normalsize\it $^a$ Center for Theoretical Physics,}\\
{\normalsize\it Laboratory for Nuclear Science and Department
 of Physics}\\
{\normalsize\it Massachusetts Institute of Technology,
 Cambridge, Massachusetts 02139}\\
{\normalsize\it $^b$ Inst. f. theoret. Physik d. Uni Wien}\\  
{\normalsize\it Boltzmanngasse 5, 1090 Wien, Austria}\\
{\normalsize\it $^c$ Departamento de F\'\i sica, 
Universidad Nacional de La Plata}\\
{\normalsize\it C.C. 67, 1900 La Plata, Argentina}\\[3mm]
(MIT-CTP-2687, ~~ hep-th/9710238 )}
\maketitle
\begin{abstract}
We compute the Schwinger terms in the energy-momentum tensor commutator
algebra from the anomalies present in Weyl-invariant and 
diffeomorphism-invariant effective actions for two dimensional massless 
scalar fields in a gravitational background. We find that the Schwinger terms 
are not sensitive to the regularization procedure and that they are 
independent of the background metric.
\end{abstract}


\section{Introduction}

The theory of a (quantized) scalar field coupled to gravity 
 has to 
follow an ad-hoc prescription: 
the functional integration over the scalar field 
$\phi$ involves the evaluation of a determinant of the Laplace operator,
which is ambiguous. 
For massless scalar fields in two-dimensional space-time the standard 
prescription implements a diffeomorphism invariant regularization that leads 
to the well known Polyakov action \cite{Polyakov} $\Gamma^{\rm P}[g^{\mu\nu}]$,
a functional of the background metric $g_{\mu\nu}$ that is indeed 
diffeomorphism invariant but has an (equally well known) anomaly with
respect to Weyl transformations. 

Recently an alternative evaluation of the theory has been given, where a 
Weyl invariant regularization has been implemented 
\cite{armenios,Jackiw,ABS,NNT}. The resulting effective action 
$\hat{\Gamma}[g^{\mu\nu}]$, while being Weyl invariant, does not remain 
invariant under general coordinate transformations, but only under those 
with unit Jacobian.

Gravitational and Weyl anomalies lead to anomalous contributions to the 
equal-time commutators of the energy-momentum tensor \cite{Teitelboim,Tomiya}
(see also \cite{lectures} for the analogous fact in current algebra). 
So the question 
arises whether these two versions of the theory lead to the same anomalous 
commutators. 
In this paper we investigate this question and find that, indeed, 
the anomalous commutators coincide in both versions of the theory and lead 
to the well known result from Conformal Field Theory \cite{Ginsparg}.
We do this calculation both for flat and curved space-time. In the latter case
of general metric the computation is done without any gauge fixing; this is 
the proper procedure because gauge fixing would be in conflict with the
Weyl-invariant 
regularization, that breaks diffeomorphism invariance.
The results, when properly interpreted, lead to the same Schwinger terms as
in the flat space-time and, therefore,
show that the Schwinger terms do not depend on the curvature.

\section{Diffeomorphism-invariant and Weyl-invariant regularizations}

First we have to fix our conventions. We use the flat Minkowskian metric 
$\eta_{ab}$ with signature $(+,-)$. The metric $g_{\mu\nu}(x)$ is related to 
the zweibein via
\beq
g_{\mu\nu}(x)=\eta_{ab} e^a_\mu(x)  e^b_\nu(x);
\eeq{1}
we also need the zweibein determinant
\beq
e(x):= {\rm det} e^a_\mu(x) = \sqrt{|{\rm det} g_{\mu\nu}(x)|}
\eeq{2}
and the inverse zweibein $E^{\mu}_a(x)$,
\beq
E^{\mu}_a(x):=\eta_{ab}g^{\mu\nu}(x) e^b_{\nu}(x).
\eeq{2ymedio}
For the curvature we use the sign convention
$R_{\mu\nu}=-\partial_{\alpha}\Gamma^{\alpha}_{\mu\nu}+\dots$,
where $R_{\mu\nu}$ is the Ricci tensor and $\Gamma^{\alpha}_{\mu\nu}$ is the 
Christoffel connection.

Weyl transformations act like
\beq
g_{\mu\nu}(x) \rightarrow \exp(2\sigma(x))g_{\mu\nu}(x), ~~~~~~~~~~~~~
e^a_\mu(x) \rightarrow \exp(\sigma(x))  e^a_\mu(x).
\eeq{3}
When the effective action 
$\Gamma$ is not invariant under Weyl transformations,
an infinitesimal change $\delta^{\rm W}_{\sigma}g_{\mu\nu}(x)= 
2\sigma(x)g_{\mu\nu}(x)$ induces a Weyl anomaly $G^{\rm W}(x)$: 
\beq
\delta^{\rm W}_{\sigma}\Gamma := \int d^2x \sigma(x) G^{\rm W}(x),
\nonumber \end{equation}
\beq
 G^{\rm W}(x)=-2 g^{\mu\nu}(x)\frac{\delta \Gamma}{\delta g^{\mu\nu}(x)}=
-e(x) g^{\mu\nu}(x)  T_{\mu\nu}(x) ,
\eeq{4}
where $T_{\mu\nu}$ is the v.e.v. of the energy momentum tensor 
$\Theta_{\mu\nu}$,
\beq
T_{\mu\nu}(x)=\langle \Theta_{\mu\nu}(x) \rangle= 
\frac{2}{e(x)}
\frac{\delta \Gamma}{\delta g^{\mu\nu}(x)}.
\eeq{5}

Under an infinitesimal coordinate transformation (diffeomor\-phism)
$\delta^{\rm D}_{\xi}x^{\mu}=-\xi^{\mu}(x)$ the metric and zweibein transform
like 
\beq
\delta^{\rm D}_{\xi} g^{\mu\nu}(x) = - D^{\mu}\xi^{\nu}(x)-D^{\nu}\xi^{\mu}(x)
, ~~~~~~~~
\delta^{\rm D}_{\xi}e^a_{\mu}(x)=\xi^{\lambda}\partial_{\lambda}e^a_{\mu}(x)
+e^a_{\lambda}(x)\partial_{\mu}\xi^{\lambda}
\eeq{6}
and a diffeomorphism anomaly is given as 
\beq
\delta^{\rm D}_{\xi}\Gamma := \int d^2x \xi^{\nu}(x) G^{\rm D}_{\nu}(x),
\nonumber \end{equation}
\beq
G^{\rm D}_{\nu}(x)=2 e(x) D^{\mu} \left( \frac{1}{e(x)} 
\frac{\delta\Gamma}{\delta 
g^{\mu\nu}(x)}\right)=
e(x)D^{\mu}T_{\mu\nu}(x).
\eeq{7}
It will be convenient later on to use covariant derivatives acting on 
the combination $eT_{\mu\nu}$, using the rule
$e D_{\alpha}= (D_{\alpha}-\Gamma^{\lambda}_{\alpha\lambda})e$.  
Thus we rewrite $G^{\rm D}_{\nu}$ as
\beq
G^{\rm D}_{\nu}(x)=(D^{\mu}-g^{\mu\rho}
\Gamma^{\lambda}_{\rho\lambda})(e(x)T_{\mu\nu}(x)).
\eeq{7a}

Further we will frequently use the following variational formulae,
\beq
\frac{\delta g^{\mu\nu}(x)}{\delta e^a_{\alpha}(y)}=
-\eta_{ac} e^c_{\lambda}(x)(g^{\mu\alpha}(x)g^{\nu\lambda}(x)+
g^{\nu\alpha}(x)g^{\mu\lambda}(x))\delta^{(2)}(x-y) ,
\eeq{8}
\beq
\frac{\delta e(x)}{\delta g^{\mu\nu}(y)}= -\frac{1}{2}e(x) g_{\mu\nu}(x)
\delta^{(2)}(x-y) ,
\eeq{9}
\beq
\frac{\delta R(x)}{\delta g^{\mu\nu}(y)}= [R_{\mu\nu}(x) +
(D_{\mu}D_{\nu}-g_{\mu\nu}\Box )_{(x)}]\delta^{(2)}(x-y) ,
\eeq{10}
where $R$ is the curvature scalar and $R_{\mu\nu}$ is the Ricci tensor.

\vspace{5mm}

The classical action of the theory reads
\beq
S=\int d^2x \frac{e(x)}{2}g^{\mu\nu}(x)
\partial_{\mu}\phi(x)\partial_{\nu}\phi(x)  .
\eeq{11}
When a diffeomorphism invariant path integration with respect to $\phi$ is 
chosen, one obtains the Polyakov effective action \cite{Polyakov}
\beq
\Gamma^{\rm P}[g^{\mu\nu}]=-\frac{1}{96\pi}\int d^2x d^2y e(x) R(x)
\Box^{-1}(x,y)
e(y) R(y),
\eeq{12}
where $\Box^{-1}(x,y)$ is the scalar symmetric Green function of the covariant
Laplacian (satisfying $\Box_{(x)}\Box^{-1}(x,y)= e^{-1}(x)\delta^{(2)}(x-y)$).
$\Gamma^{\rm P}$ is diffeomorphism invariant, 
\beq
G^{\rm D}_{\nu}(x)= 0,
\nonumber \end{equation}
and posseses the well known Weyl anomaly (for a comprehensive review, see for 
instance \cite{Duff} and references therein),
\beq
G^{\rm W}(x) =-\frac{1}{24 \pi}e(x) R(x).
\eeq{13}

The alternative, Weyl invariant evaluation that was discussed in 
\cite{armenios,Jackiw,ABS,NNT} relies on the observation that the classical 
action (\ref{11}) depends only on the Weyl invariant quantity 
$\gamma^{\mu\nu}$, where
\beq
\gamma^{\mu\nu}(x)=e(x)g^{\mu\nu}(x), ~~~~~~~~~
\gamma_{\mu\nu}(x)=\frac{1}{e(x)}g_{\mu\nu}(x).
\eeq{14}
As the breaking of the classical Weyl invariance 
in Polyakov's path integration may be traced back to a 
diffeomorphism-invariant and Weyl non-invariant normalization for 
the path integral measure,
\beq
\int \D \phi \exp(i\int d^2x e(x)\phi^2(x))=1,
\eeq{15}
the Weyl invariant evaluation can be achieved by choosing instead
\beq
\int \D \phi \exp(i\int d^2x \phi^2(x))=1.
\eeq{16}
This leads to a Weyl-invariant effective action $\hat{\Gamma}[g^{\mu\nu}]$
which depends on $g^{\mu\nu}(x)$ only through the combination 
$\gamma^{\mu\nu}$. By construction the two effective actions 
$\Gamma^{\rm P}$ and $\hat{\Gamma}$ coincide for metrics 
with unit determinant, therefore
\beq
\hat{\Gamma}[g^{\mu\nu}] \equiv \Gamma^{\rm P}[\gamma^{\mu\nu}]=
-\frac{1}{96\pi}\int d^2x d^2y \hat{R}(x)\Box^{-1}(x,y) 
\hat{R}(y),
\eeq{17}
where $\hat{R}(x)$ is the curvature scalar evaluated from $\gamma^{\mu\nu}$
(notice that $\hat{R}(x)$ is not a true scalar). 

$\hat{\Gamma}$ is Weyl-invariant, but it acquires an anomaly under coordinate
transformations with Jacobian not equal to unity. This anomaly may actually
be easily computed from the Weyl anomaly of the Polyakov action. The v.e.v.
of the energy-momentum tensor computed from $\hat{\Gamma}$ is
\beqn
\hat{T}_{\mu\nu}(x) & = & \frac{2}{e(x)}
\frac{\delta \hat{\Gamma}}{\delta g^{\mu\nu}(x)}=
2\frac{\delta \hat{\Gamma}}{\delta \gamma^{\mu\nu}(x)}-\gamma_{\mu\nu}
\gamma^{\alpha\beta}\frac{\delta \hat{\Gamma}}{\delta \gamma^{\alpha\beta}(x)}
\nonumber\\
&=& 
 T^{\rm P}_{\mu\nu}(\gamma) -\frac{1}{2}\gamma_{\mu\nu}\gamma^{\alpha\beta}
 T^{\rm P}_{\alpha\beta}(\gamma).
\eeqn{18}
Here $T^{\rm P}_{\mu\nu}(\gamma)$ is the energy-momentum tensor 
$T^{\rm P}_{\mu\nu}$, 
as computed from the Polyakov action,
evaluated at $g^{\mu\nu}=\gamma^{\mu\nu}$. Obviously, there 
is no Weyl anomaly, $g^{\mu\nu}\hat{T}_{\mu\nu}=0$.

In order to evaluate the diffeomorphism anomaly we need the identity
$D_{\mu}(g^{\mu\nu}\hat{T}_{\nu\alpha})= 
\frac{1}{e}\hat{D}_{\mu} (\gamma^{\mu\nu}\hat{T}_{\nu\alpha})$, 
which may be easily proven by using the tracelessness and symmetry of 
$\hat{T}_{\mu\nu}$ (here $\hat{D}_{\mu}$ is the covariant derivative for 
the metric $\gamma^{\mu\nu}$). We then find for the diffeomorphism anomaly
\beqn
\hat{G}^{\rm D}_{\alpha}&=& e D_{\mu}(g^{\mu\nu}\hat{T}_{\nu\alpha}) 
\nonumber\\
&=& \hat{D}_{\mu}\left(\gamma^{\mu\nu}T^{\rm P}_{\nu\alpha}(\gamma) 
-\frac{1}{2}\gamma^{\mu\nu}\gamma_{\nu\alpha}\gamma^{\beta\delta}
 T^{\rm P}_{\beta\delta}(\gamma)\right)
\nonumber\\
&=& -\frac{1}{2} \hat{D}_{\alpha}(\gamma^{\beta\delta}
T^{\rm P}_{\beta\delta}(\gamma) )
\nonumber\\
&=& -\frac{1}{48\pi}\partial_{\alpha}\hat{R}.
\eeqn{19}
Here we have used the vanishing of the diffeomorphism anomaly for 
$\Gamma^{\rm P}$
and the fact that $\hat{D}_{\alpha}$ reduces to the ordinary derivative 
on scalars.
The anomaly is a pure divergence because only the symmetry with respect to 
transformations with non-unit Jacobian is broken (see \cite{Jackiw}).

\section{Schwinger terms}

In this section we want to relate the anomalies of the previous section to the
equal-time commutators (ETCs) of the energy-momentum tensor, both in flat
and curved space-time. Here we will 
follow a
method that was developed in \cite{Ebner} and used there for the calculation
of ETCs in the flat space-time limit. We want to find the 
Schwinger terms in the general case of a non flat space-time, too, which makes 
things slightly more complicated. We choose the hypersurface $x^0=0$ as a 
quantization surface. For ETCs we write
\beq
\delta(x^0-y^0)[e(x)\Theta^{\mu}_a(x), e(y)\Theta^{\nu}_b(y)] =
\Theta^{\mu\nu}_{ab}(x,y) + S^{\mu\nu}_{ab}(x,y),
\eeq{20}
where we have used the zweibein formalism in order  to conform with 
\cite{Ebner} (i.e. $\mu$, $\nu$ are space-time indices whereas $a$, $b$ are 
Lorentz indices). In eq.\ (\ref{20}) $\Theta^{\mu\nu}_{ab}$ is the canonical
part, depending again on the regularized energy-momentum operators 
$\Theta^{\mu}_a(x)$, 
whereas $S^{\mu\nu}_{ab}$ are $c$-numbers (the Schwinger terms). In the flat
case regularization means just normal ordering, and therefore the v.e.v.\ of 
eq.\ (\ref{20}) arises only from  $S^{\mu\nu}_{ab}$ in the r.h.s. In the 
general case this is no longer true \cite{Bos} but our knowledge of the flat 
case will still enable us to identify the individual pieces. 

In the flat case
it is well known that the canonical part is proportional to the first spatial
derivative of the delta function, e.g. $\Theta^{00}_{01}(x,y) \sim
i(\Theta^0_0(x)+\Theta^0_0(y))\delta(x^0-y^0)\delta'(x^1-y^1)$, whereas 
the Schwinger term is proportional to a triple spatial derivative, 
$S^{00}_{01}(x,y) \sim c \delta(x^0-y^0)\delta'''(x^1-y^1)$ 
($c$ is a constant).

In the general case both the expression for the classical energy-momentum 
tensor (see 
(\ref{11})) and the regularization will introduce a dependence on the metric 
and its derivatives in eq.\ (\ref{20}). However, we will assume that the
number of derivatives on the delta function remains unchanged, i.e.\ we will 
continue to identify the $\delta'''$ piece of the v.e.v.\ of eq.\ (\ref{20})
with the Schwinger term. By treating the deviation from the flat space-time
action (\ref{11}) as interaction, $S_{\rm I}=S[g^{\mu\nu}]- S[\eta^{\mu\nu}]$,
$\Gamma=-i \ln <0|T^* \exp iS_{\rm I}|0> = -i \ln {\cal Z}=-i \ln <out|in>$
we find for the two point function
\beqn
-i \frac{\delta^2\Gamma}{\delta e^a_{\mu}(x) e^b_{\nu}(y)}&=&
<out|T^*(e(x) \Theta^{\mu}_a(x) e(y)\Theta^{\nu}_b(y))|in>
\nonumber\\
&-& <out|e(x) \Theta^{\mu}_a(x)|in> 
<out|e(y) \Theta^{\nu}_b(y)|in> 
\nonumber \\
&+& <out|\frac{1}{i}\frac{\delta((e(x) \Theta^{\mu}_a(x))}
{\delta e^b_{\nu}(y)}|in> 
\nonumber \\
&:=& T^{\mu\nu}_{ab}(x,y) + \Omega^{\mu\nu}_{ab}(x,y),
\eeqn{21}
where $T^{\mu\nu}_{ab}(x,y)$ is the connected, time-ordered two-point function
\beqn
 T^{\mu\nu}_{ab}(x,y)&=&<out|T (e(x) \Theta^{\mu}_a(x) e(y)
\Theta^{\nu}_b(y))|in> \nonumber \\
&-& <out|e(x) \Theta^{\mu}_a(x)|in> 
<out|e(y) \Theta^{\nu}_b(y)|in> 
\eeqn{22}
and $\Omega^{\mu\nu}_{ab}$ contains the remaining pieces and is local
(i.e.\ proportional to $\delta(x-y)$ and derivatives thereof).

Now we want to relate this two-point function to functional derivatives of the
anomalies in eqs.\ (\ref{4},\ref{7}). Defining these functional derivatives as 
\beq
I^{\alpha}_{ab}(x,y):= -i E_a^{\mu} 
\frac{\delta G^{\rm D}_{\mu}(x)}{\delta e^b_{\alpha}(y)}, 
\eeq{23}
\beq
\Pi^{\alpha}_b(x,y):=-i\frac{\delta G^{\rm W}(x)}{\delta e^b_{\alpha}(y)},
\eeq{24}
we find the relations
\beqn
-I^{\alpha}_{ab}(x,y)+A^{\alpha}_{ab}(x,y)&=&
(D_{\rho}-\Gamma^{\lambda}_{\rho\lambda})_{(x)}(T^{\rho\alpha}_{ab}(x,y)
+  \Omega^{\rho\alpha}_{ab}(x,y))
\nonumber \\
&=& S^{0\alpha}_{ab}(x,y) + (D_{\rho}-\Gamma^{\lambda}_{\rho\lambda})_{(x)}
\Omega^{\rho\alpha}_{ab}(x,y)
\eeqn{25}
and
\beq
\Pi^{\alpha}_b(x,y) + B^{\alpha}_b(x,y) = e^a_{\mu}(x)
\Omega^{\mu\alpha}_{ab}(x,y) .
\eeq{26}
Here $A^{\alpha}_{ab}(x,y)$ and $ B^{\alpha}_b(x,y)$ stem from variations 
of the anomalies (\ref{4},\ref{7}) that do not vary the one-point function 
$e(x)T_{\mu\nu}(x)$ (e.g.\  $B^{\alpha}_b(x)(x,y) =
-\left( \frac{\delta g^{\mu\nu}(x)}{\delta e^b_{\alpha}(y)}\right) 
e(x)T_{\mu\nu}(x)$). They produce $\delta$ functions and first derivatives 
thereof and vanish in the flat limit. 
They are unimportant in the sequel. Further, we have 
assumed in eqs.\ (\ref{25},\ref{26}) that the anomalies of the 
Heisenberg operators 
$\Theta ^a_{\mu}$ are themselves $c$-numbers. Under this assumption the 
anomalies do not contribute to the connected two-point function, e.g.\
$<T\left((D^{\mu}\Theta_{\mu\nu}(x))\Theta_{\alpha\beta}(y)\right)>_c =0$. 
(Here we slightly differ in the conventions from \cite{Ebner}. They treat the 
operator $\Theta _a^{\mu}(x)$ as an interaction picture operator and, 
therefore, obtain additional commutators $[\Theta_a^0(x), L_{\rm I}(x^0)]$ 
in their relations.)

As we use the 
zweibein formalism, we need the corresponding equation for the Lorentz 
anomaly, even though the latter vanishes in both regularizations of 
our theory. Under infinitesimal Lorentz transformations the
zweibein changes as
\beq
\delta^{\rm L}_{\alpha} e^a_{\mu}= - \alpha^a_b e^b_{\mu},
\eeq{27}
inducing a variation of the effective action
\beq
\delta^{\rm L}_{\alpha} \Gamma := \int d^2x \alpha^{ab}G^{\rm L}_{ab}(x)
\eeq{28}
where
\beq
G^{\rm L}_{ab}(x)= -\frac{1}{2}(\eta_{ac}e^c_{\mu}\frac{\delta}
{\delta e^b_{\mu}}
-\eta_{bc}e^c_{\mu}\frac{\delta}{\delta e^a_{\mu}}) \Gamma.
\eeq{29}
Then, defining
\beq
L^{\alpha}_{cab}(x,y) :=-i \frac{\delta G^{\rm L}_{ab}(x)}{\delta 
e^c_{\alpha}(y)}
\eeq{30}
we find a further set of equations
\beq
L^{\alpha}_{cab}(x,y) + C^{\alpha}_{cab}(x,y) = 
\eta_{cd} e^d_{\mu}\Omega^{\mu\alpha}_{ab}(x,y) - 
\eta_{ad} e^d_{\mu}\Omega^{\mu\alpha}_{cb}(x,y)
\eeq{31}
(where $ C^{\alpha}_{cab}$ is irrelevant, analogous to the above $A$ and $B$).

Next we need the explicit expressions for the functional derivatives of the 
anomalies ($L^{\alpha}_{cab}$ being zero in both cases of interest). 
For the Polyakov action $\Gamma^{\rm P}$ we have $G^{\rm D}=0$ and 
\beqn
\frac{\delta G^{\rm W}(x)}{\delta e^b_{\alpha}(y)}&=& -\frac{1}{24\pi}
\int d^2z
\frac{\delta (e(x)R(x))}{\delta g^{\mu\nu}(z)}
\frac{\delta g^{\mu\nu}(z)}{\delta e^b_{\alpha}(y)} \label{32} \\
&=& \frac{1}{24\pi} \eta_{bc}e^c_{\lambda}(y)(g^{\mu\alpha}g^{\nu\lambda}
+g^{\mu\lambda}g^{\nu\alpha})_{(y)} (D_{\mu}D_{\nu}-g_{\mu\nu}\Box)_{(x)}
\delta^{(2)}(x-y),
\nonumber
\end{eqnarray} 
whereas for the Weyl-invariantly regularized effective action $\hat\Gamma$
we find $\hat{G}^{\rm W}=0$ and  
\beqn
\frac{\delta\hat{G}^{\rm D}_{\lambda}(x)}{\delta e^b_{\alpha}(y)}&=& 
-\frac{1}{48\pi}
\partial_{\lambda}^x \int d^2z d^2z'
\frac{\delta \hat{R}(x)}{\delta \gamma^{\rho\sigma}(z)}
\frac{\delta \gamma^{\rho\sigma}(z)}{\delta g^{\beta\delta}(z')}
\frac{\delta g^{\beta\delta}(z')}{\delta e^b_{\alpha}(y)}
\nonumber\\
&=& 
\frac{e(y)}{48\pi}(\delta^{\rho}_{\mu}\delta^{\sigma}_{\nu}-\frac{1}{2}
g_{\mu\nu}g^{\rho\sigma})_{(y)}\eta_{bc}e^c_{\epsilon}(y)
(g^{\beta\alpha}g^{\delta\epsilon}+g^{\delta\alpha}g^{\beta\epsilon})_{(y)}
\nonumber\\
& & \times \partial_{\lambda}^x (\hat{D}_{\rho}\hat{D}_{\sigma}-
\gamma_{\rho\sigma}\hat{\Box})_{(x)}
\delta^{(2)}(x-y).
\eeqn{33}

Now the procedure of \cite{Ebner} for evaluating the Schwinger terms 
$S^{0\alpha}_{ab}$ consists in expanding all the local functions of 
eqs.\ (\ref{25},\ref{26},\ref{31}) into derivatives of $
\delta$ functions, e.g.\
\beq
I^{\alpha}_{ab}(x,y)=\sum_{n,k}I^{\alpha(k,n-k)}_{ab}(x)
\partial_0^k\partial_1^{n-k}\delta^{(2)}(x-y).
\eeq{34}
The index $k=0,\cdots ,n$ counts the number of time derivatives, while $n-k$
counts space derivatives. 
In particular, $S^{0\alpha}_{ab}(x,y)$ has only 
spatial derivatives of $\delta$ functions,
\beq
S^{0\alpha}_{ab}(x,y)=\sum_{n}S^{0\alpha(n)}_{ab}
\partial_1^{n}\delta^{(2)}(x-y).
\eeq{34a}
Thus,
one obtains a system of linear equations for the unknown coefficient functions
$S^{0\alpha(n)}_{ab}$ and $\Omega^{\mu\alpha(k,n-k)}_{ab}$.

First let us briefly review the flat space-time computation that was done in
\cite{Ebner} (they used it for chiral fermions, too, where diffeomorphism and  
Weyl anomalies are present). 
In this case all derivatives only act on
the $\delta$ functions. Therefore the explicit 
expression analogous to (\ref{33}) for 
$I^{\alpha}_{ab}$ contains only terms with three derivatives, and the 
corresponding expression (\ref{32})
for  $\Pi^{\alpha}_{a}$ only terms with two derivatives. 
Further, the covariant 
derivative in eq.\ (\ref{25}) turns into an ordinary derivative. As a 
consequence, the resulting system of equations may be solved separately for
each fixed number of derivatives ($n$ derivatives for $I$, $S$ and $n-1$
derivatives for $\Pi$, $\Omega$); for each fixed $n$ the number of 
unknowns $S^{0\alpha(n)}_{ab}$ and $\Omega^{\mu\alpha(k,n-k)}_{ab}$
equals the number of equations. As only  $\Pi^{\alpha(k,2-k)}_{a}$ and
$I^{\alpha(k,3-k)}_{ab}$ are non-zero, one finds  a non-zero result only for
$S^{0\alpha(3)}_{ab}$,  $\Omega^{\mu\alpha(k,2-k)}_{ab}$ 
(even in the non-flat case, we will only consider 
the coefficient of the triple derivative of the Schwinger term, 
therefore we drop the superscript $(3)$).
Eliminating the 
$\Omega$s, one arrives at the flat space result
\beq
S^{0\alpha}_{0b}=-I^{\alpha(0,3)}_{0b}-I^{\alpha(1,2)}_{1b}-
I^{\alpha(2,1)}_{0b}-I^{\alpha(3,0)}_{1b}-\Pi^{\alpha(1,1)}_b,
\eeq{35}
\beq
S^{0\alpha}_{1b}=-I^{\alpha(0,3)}_{1b}-I^{\alpha(1,2)}_{0b}-
I^{\alpha(2,1)}_{1b}-I^{\alpha(3,0)}_{0b}-\Pi^{\alpha(0,2)}_b-
\Pi^{\alpha(2,0)}_b.
\eeq{36}
These equations we have to evaluate for the two versions $\Gamma^{\rm P}$ and
$\hat\Gamma$ of our theory in the flat limit. In the first case only 
$\Pi^{\alpha}_b$ are non-zero, in the second case only $\hat{I}^{\alpha}_{ab}$.
Both versions lead to the same Schwinger terms,
\beq
S^{00}_{00}=S^{00}_{11}=0,
\eeq{37}
\beq
S^{00}_{01}=S^{00}_{10}=\frac{i}{12\pi}.
\eeq{38}
For the Weyl anomaly this result was in fact already computed in \cite{Ebner}
(we differ in signs because of different metric and curvature conventions).
For the diffeomorphism anomaly we find the same result, showing that the 
Schwinger terms are not sensitive to the regularization prescription.

\vspace{5mm}

Next we want to discuss the case of general metric. In this case one has 
covariant derivatives in eqs.\ (\ref{25},\ref{32},\ref{33}), 
and therefore the system of 
equations (\ref{25},\ref{26},\ref{31}) 
mixes different number of derivatives. However,
$I^{\alpha}_{ab}$ and $\Pi^{\alpha}_{b}$ still contain at most three and two 
derivatives, respectively, acting on $\delta$ functions. 
If one also assumes that $\Omega^{\alpha\mu}_{ab}$
contains at most two derivatives (which is a very reasonable assumption, as 
all diagrams contributing to 
$<T(e(x)\Theta^{a}_{\mu}(x)e(y)\Theta^{b}_{\nu}(y))>$
are at most quadratically divergent), it still holds that the subsystem of 
equations containing the maximal number of derivatives (three for $I$, $S$
and two for $\Pi$, $\Omega$) may be solved separately.

This system of equations is a little bit more complicated and leads 
again to the 
same solution for both the Weyl anomaly of $\Gamma^{\rm P}$ or the 
diffeomorphism
anomaly of $\hat\Gamma$. The coefficients of $\partial_1^3 \delta^{(2)}(x-y)$ 
in the Schwinger terms read
\beqn
S^{00}_{00}=S^{00}_{11} &=& -\frac{i}{6\pi}\frac{e^0_1 e^1_1}{(g_{11})^2},
\nonumber \\
S^{00}_{10}=S^{00}_{01} &=& \frac{i}{12\pi}\frac{(e^0_1)^2 +  (e^1_1)^2}
{(g_{11})^2},
\eeqn{39}
and (defining $\kappa=\frac{i}{12\pi e (g_{11})^3}$)
\beqn
S^{01}_{00} &=& \kappa (-e^0_0 e^0_1 g_{01} g_{11} -e^0_0 e^1_1 e g_{11}
+ (e^0_1)^2((g_{01})^2+e^2) + 2 e^0_1 e^1_1 e g_{01}) ,
\nonumber\\
S^{01}_{01} &=& \kappa (e^0_1 e^1_0 g_{01} g_{11} +e^1_0 e^1_1 e g_{11}
- e^0_1 e^1_1 ((g_{01})^2+e^2) - 2 ( e^1_1)^2 e g_{01}), 
\nonumber\\
S^{01}_{10} &=& \kappa (e^0_0 e^1_1 g_{01} g_{11} +e^0_0 e^0_1 e g_{11}
- e^0_1 e^1_1 ((g_{01})^2+e^2) - 2 ( e^0_1)^2 e g_{01}) ,
\nonumber\\
S^{01}_{11} &=& \kappa (-e^1_0 e^1_1 g_{01} g_{11} -e^0_1 e^1_0 e g_{11}
+ (e^1_1)^2((g_{01})^2+e^2) + 2 e^0_1 e^1_1 e g_{01}) . \nonumber \\
~~
\eeqn{40}
Although some components look rather ugly, this result is precisely what
one expects, as we want to discuss now. 

Let us transform 
$S^{\mu\alpha}_{ab}$ to pure space-time indices via
\beq
S^{\mu'\alpha'}_{ab}= E_a^{\nu}E_b^{\beta} g^{\mu'\mu} g^{\alpha'\alpha}
S_{\mu\nu\alpha\beta}.
\eeq{41}
Notice that we cannot invert this relation because we do not know all the 
components of $S^{\mu\alpha}_{ab}$. However, due to the symmetries
$S_{\mu\nu\alpha\beta}=S_{\nu\mu\alpha\beta}=S_{\alpha\beta\mu\nu}$,
$S_{\mu\nu\alpha\beta}$ actually consists of six independent components. 
The expressions (\ref{39},\ref{40}) for $S^{\mu'\alpha'}_{ab}$ lead to five 
independent equations for $S_{\mu\nu\alpha\beta}$. Therefore we are able to 
express all components of $S_{\mu\nu\alpha\beta}$ in terms of one
unknown function $\Lambda$, where the form of $\Lambda$ is restricted by the
requirement that all $S_{\mu\nu\alpha\beta}$ tend to their well known 
Minkowski space version in the flat limit. We obtain
\beqn
S_{0000}&=& \frac{4i e^3 g_{00} g_{01}}{12 \pi (g_{11})^3} -
\frac{8i e^3 ( g_{01})^3}{12 \pi (g_{11})^4} 
+\frac{(g_{01})^5}{(g_{11})^4} \Lambda  ,
\nonumber \\
S_{0001}&=& \frac{i e^3 g_{00}}{12 \pi (g_{11})^2} -
\frac{4i e^3 ( g_{01})^2}{12 \pi (g_{11})^3}
+\frac{(g_{01})^4}{(g_{11})^3} \Lambda ,
\nonumber \\
S_{0101}=S_{0011} &=& -\frac{2i e^3 g_{01}}{12\pi (g_{11})^2}+
\frac{(g_{01})^3}{(g_{11})^2} \Lambda  ,
\nonumber \\
S_{0111} &=& -\frac{i e^3 }{12 \pi g_{11}}+
\frac{(g_{01})^2}{g_{11}} \Lambda ,
\nonumber \\
S_{1111}&=& g_{01}\Lambda ,
\eeqn{42}
where $\Lambda$ may be non-zero (but finite) in the flat limit.

For a proper interpretation of this result we need some basic facts about 
canonical quantization in curved space-time. We chose the hypersurface
$x^0={\rm const}$ as a quantization surface. The direction of the (arbitrarily 
chosen) time coordinate is not an intrinsic property of this surface, and, 
therefore, time components of tensors are not invariant 
under coordinate transformations that do not change the coordinates on the 
hypersurface. Instead one has to
choose the projection of the time components onto the timelike vector
$l^{\mu}$ orthogonal to the surface, e.g.\ ($T_{\mu\nu}$ is a general tensor, 
$i$ is the space index)
\beq
T_{\mu\nu} \rightarrow T_{ij}, ~~l^{\mu}T_{\mu j}, ~~l^{\nu}T_{i \nu},
~~l^{\mu}l^{\nu}T_{\mu\nu}
\eeq{43}
(see e.g.\ \cite{Dirac}).
The vector $l^{\mu}$ is given by
\beq
l^{\mu}= e g^{0\mu}.
\eeq{44}
Here we chose the normalization $l^{\mu}l_{\mu}=-g_{11}$, which is the proper 
normalization in order to obtain the correct commutator algebra on the 
quantization surface, see e.g.\ \cite{Teitelboim,Teitelboim2}
(this normalization corresponds to the requirement that $l^{\mu}$ is a 
vector, not a vector density: for a general tangent vector $b_1^{\mu}$ to
the hypersurface, the orthogonal covector $l_{\mu}$ is
$l_{\mu}=\bar{\epsilon}_{\mu\nu}b_1^{\nu}$, where $\bar{\epsilon}_{\mu\nu}=
e^a_{\mu}e^b_{\nu} \epsilon_{ab}= e\epsilon_{\mu\nu}$ is a tensor. For our 
specific choice $b_1^{\mu}=\delta_1^{\mu}$ one finds precisely (\ref{44}) for 
$l_{\mu}$).
Further we should remember that $S^{\mu\nu}_{ab}$ was defined as the 
commutator of $[e(x)\Theta^{\mu}_a(x), e(y)\Theta^{\nu}_b(y)]$ 
(see eq.\ (\ref{20})), i.e.\ to obtain the commutators of the $\Theta^{\mu}_a$
themselves we still have to divide by $e^2$. Doing so, and performing the
projections, we recover precisely the central extension of the Virasoro algebra
\cite{Virasoro}
\beq
l^{\mu}l^{\nu}l^{\alpha}l^{\beta}S_{\mu\nu\alpha\beta}=
l^{\mu}l^{\nu}S_{\mu\nu 11}= 
l^{\mu}l^{\alpha}S_{\mu 1 \alpha 1}=0
\eeq{45}
\beq
e^{-2}(x)l^{\mu}l^{\nu}l^{\alpha}S_{\mu\nu\alpha 1}= 
e^{-2}(x)l^{\mu}S_{\mu 111}= 
\frac{i}{12 \pi}
\eeq{46}
and the arbitrary function $\Lambda$ cancels out in all expressions (\ref{45},
\ref{46}). The pure space component $S_{1111}= g_{01}\Lambda$, 
which is not related to any symmetry generator, remains undetermined by our 
procedure. 

\section{Conclusions}

We have analyzed the anomalous Schwinger terms in the equal-time 
energy-momentum tensor algebra in two different  regularizations of
2-d scalar field theory in a curved background.

The usual computations make use of the conformal gauge, which is of course
appropriate for the diffeomorphism-invariant regularization. Once the metric 
is set to its conformally flat form, all the machinery of Conformal Field 
Theory can be applied essentially as in flat space-time \cite{FMS}.
In contrast, the gauge fixing can not be performed in the Weyl-invariant 
version of the theory. In order to compare both regularizations one then needs
a more general framework, in which no gauge fixing is made at any step.

In this framework 
we have achieved a two-fold result. On the one hand, we have shown that
the energy-momentum operators continue to obey the Virasoro algebra in the
case of a general metric, {\em without using any gauge fixing} 
for the computation.  
On the other hand, we have proven that both versions of the theory,
eq.\ (\ref{12}) and eq.\ (\ref{17}), obey the 
same commutation relations, regardless of the symmetries broken by the 
regularization procedures.

\underline{Acknowledgements}: The authors are grateful to Prof.\ Roman Jackiw
for suggesting the problem and for guidance throughout the work. C.A.\ is 
supported by a Schr\"odinger stipendium of the Austrian FWF.
G.L.R.\ is partially supported by CONICET, Argentina. 
This work is supported in part by funds provided by
the U.S.\ Department of Energy (D.O.E.) under cooperative research
agreement \#DF-FC02-94ER40818.

\end{document}